\documentclass[
    aps,
    prd,
    reprint,
    amsmath,
    amssymb,
    nofootinbib,
    floatfix
]{revtex4-2}

\usepackage[T1]{fontenc}
\usepackage{lmodern}
\usepackage{microtype}

\usepackage{mathtools}
\usepackage{bm}

\usepackage{graphicx}
\graphicspath{{figures/}}

\usepackage{booktabs}
\usepackage{placeins}

\usepackage{xcolor}
\usepackage[
    colorlinks=true,
    linkcolor=blue!65!black,
    citecolor=blue!65!black,
    urlcolor=blue!65!black
]{hyperref}
\usepackage[capitalise]{cleveref}
\crefname{section}{Sec.}{Secs.}
\Crefname{section}{Section}{Sections}
\crefname{subsection}{Sec.}{Secs.}
\Crefname{subsection}{Section}{Sections}
\crefname{appendix}{Appendix}{Appendices}
\crefname{table}{Table}{Tables}

\newcommand{\dd}{\mathrm{d}}
\newcommand{\pd}{\partial}
\newcommand{\cd}{\nabla}
\newcommand{\vphi}{v_{\phi}}
\newcommand{\Omcl}{\Omega_{\mathrm{cl}}}
\newcommand{\Omz}{\Omega_{0}}
\newcommand{\Omq}{\Omega_{q}}
\newcommand{\Kcl}{K_{\mathrm{cl}}}
\newcommand{\Br}{\mathrm{Br}}

\begin{document}

\title{Special-relativistic Taylor--Couette flow: Viscous shear, radial heat
       conduction, and energy conservation}

\author{Ani Gelekva}
\email{aghel22@freeuni.edu.ge}
\affiliation{School of Physics, Free University of Tbilisi, Tbilisi 0183, Georgia}
\affiliation{Ilia State University, Tbilisi 0162, Georgia}

\date{\today}

\begin{abstract}
The stationary laminar flow of a viscous relativistic fluid between two
infinitely long, coaxial rotating cylinders is investigated in flat spacetime.
When the zero-heat-flux approximation is extended to this geometry, the angular
velocity $\Omega(r)$ obeys a nonlinear equation in which the shear enters with
the factor $\gamma^{3}$, where $\gamma=(1-r^{2}\Omega^{2})^{-1/2}$. However,
the energy equation of this model is violated by the local viscous heating
rate divided by $\gamma$, so that a stationary state would require a local
heat sink. The problem is therefore formulated in the Eckart frame with a
radial heat flux. Since heat carries momentum, the angular-momentum balance
then contains $\gamma$ instead of $\gamma^{3}$, together with a term
proportional to the conserved energy flux. An exact integral of the
temperature equation relates this flux to the wall temperatures and
demonstrates that the flow is independent of the thermal conductivity
$\kappa$ when the ratio of the wall temperatures is determined by the
centripetal term of Eckart's law. This condition reduces to equal plate
temperatures in planar flow and to the Tolman--Ehrenfest relation in rigid
rotation, whereas for other wall temperatures the torque depends on $\kappa$.
For a radius ratio of 2, a resting outer cylinder, and an inner wall speed of
$0.9c$ ($0.99c$), the torque exceeds its classical value by a factor of $4.89$
($134$) in the zero-heat-flux model and of $2.71$ ($23.2$) in the
$\kappa$-independent heat-conducting solution. Both formulations reduce to the
classical profile at small speeds and to the corresponding planar solutions in
a narrow gap.
\end{abstract}

\maketitle

\section{Introduction}
\label{sec:intro}

Taylor--Couette flow, the laminar motion of a viscous fluid confined between
two coaxial rotating cylinders \cite{LandauLifshitz1987}, is one of the
simplest configurations in which differential rotation can be studied in
isolation. In the classical incompressible regime the azimuthal velocity has
the form $\vphi(r)=Ar+B/r$, and the flow serves both as a laboratory system and
as an idealized reference problem for rotating shear flows in astrophysics
\cite{JiGoodman2023}. The Navier--Stokes description, however, applies only
when the fluid speed is much smaller than the speed of light; at relativistic
speeds it must be replaced by a covariant formulation based on the
conservation of the stress-energy tensor, $\cd_{\alpha}T^{\alpha\beta}=0$.

Stationary relativistic shear flows between parallel plates were analyzed by
Rogava \cite{Rogava1996} for a fluid of constant viscosity and vanishing heat
flux, and by Granik and Chapline \cite{GranikChapline1996} with Eckart heat
conduction. More recently, Gavassino \emph{et al.}\ \cite{Gavassino2026}
have shown that, in planar Couette flow, setting the heat flux to zero leaves
the energy equation $\cd_{\alpha}T^{\alpha t}=0$ unsatisfied. Since viscous
shear continuously converts mechanical work into heat, a stationary state
requires this heat to leave through the boundaries; in relativity, the heat
flux also carries momentum and therefore enters the momentum balance directly
\cite{Gavassino2026}. Although the planar problem provides the natural
reference, the cylindrical problem cannot be obtained from it by a change of
coordinates. In particular, the shear is nonuniform across the gap, and the
fluid elements have a nonzero centripetal four-acceleration, which enters
Eckart's heat law and couples the temperature profile to the rotation
\cite{Eckart1940,GavassinoAntonelli2025}. These differences have a direct
consequence for the heat-conducting problem. In the planar case the flow is
independent of the conductivity for equal plate temperatures
\cite{Gavassino2026}, whereas in the cylindrical case the corresponding
condition involves the inertial term and therefore depends on the solution
itself; any departure from it enters the angular-momentum flux through the
energy flux. Neither effect can be obtained by adapting the planar solution.

In the present work, the special-relativistic Taylor--Couette base flow is
examined by comparing the zero-heat-flux approximation with a heat-conducting
formulation in the Eckart frame \cite{Eckart1940}. The principal results are
the following:
\begin{enumerate}
    \item[(i)] For the zero-heat-flux model, the residual of the energy
    equation is determined exactly; it equals the local viscous heating rate
    divided by $\gamma$ [\cref{eq:energy_violation}].
    \item[(ii)] For the heat-conducting model, the energy and
    angular-momentum equations have two exact first integrals. A reduced
    angular-momentum equation follows from them, in which $\gamma^{3}$ is
    replaced by $\gamma$ and a term proportional to the energy flux appears
    (\cref{subsec:heat_coupled}), and an integral of the temperature equation
    establishes an exact relation between the energy flux, the
    angular-momentum flux, the profile, and the wall temperatures
    (\cref{subsec:thermal}).
    \item[(iii)] It follows that the flow is independent of $\kappa$ for walls
    whose temperature ratio is fixed by the centripetal term of Eckart's law
    [\cref{eq:balanced}]. This condition generalizes both the equal plate
    temperatures of the planar problem and the Tolman--Ehrenfest relation of
    rigid rotation, and the same solution is obtained in the limit
    $\kappa\to0$ at fixed wall temperatures. For other wall temperatures, the
    torque depends on $\kappa$ already at order $v^{2}/c^{2}$, through a single
    mismatch parameter that, together with the direction of rotation,
    determines whether the torque increases or decreases with $\kappa$
    (\cref{subsec:thermal,subsec:limits}).
    \item[(iv)] Numerical solutions of both models are compared with the
    classical profile, with the leading $O(v^{2}/c^{2})$ corrections, and with
    the planar solutions in a narrow gap, and benchmark values are provided for
    use as a test problem (\cref{sec:numerics,app:numerics}); the code is
    publicly available \cite{GelekvaCode2026}.
\end{enumerate}
To the author's knowledge, the special-relativistic Taylor--Couette problem
with first-order heat conduction has not been treated before.

The model is restricted to flat spacetime and excludes gravity, magnetic
fields, turbulence, end effects of finite cylinders, and second-order
(causal) relaxation terms. Although these restrictions exclude most realistic
applications, they isolate the effects of special-relativistic kinematics and
first-order radial heat conduction on the laminar base flow. The metric
signature $(-,+,+,+)$ and units with $c=1$ are used throughout, with $c$
restored in nonrelativistic limits, together with cylindrical coordinates
$x^{\mu}=(t,r,\phi,z)$; the general formalism is described in
Ref.~\cite{RezzollaZanotti2013}. The symbol $T$ denotes the temperature and
$T^{\alpha\beta}$ the stress-energy tensor.

\section{Classical Taylor--Couette baseline}
\label{sec:classical}

Consider a classical incompressible fluid of shear viscosity $\eta$ between two
infinitely long coaxial cylinders of radii $R_{1}<R_{2}$ rotating with constant angular
velocities $\Omega_{1}$ and $\Omega_{2}$ \cite{LandauLifshitz1987}. For
stationary, axisymmetric, purely azimuthal laminar motion, $v_{r}=v_{z}=0$ and
$\vphi(r)=r\Omega(r)$. The azimuthal component of the Navier--Stokes equation
then gives
\begin{equation}
    \Omcl(r) = A + \frac{B}{r^{2}},
    \qquad
    v_{\phi,\mathrm{cl}}(r) = A r + \frac{B}{r},
    \label{eq:classical_profile}
\end{equation}
and the no-slip conditions $\Omega(R_{1})=\Omega_{1}$ and
$\Omega(R_{2})=\Omega_{2}$ fix the constants:
\begin{equation}
    A = \frac{\Omega_{2} R_{2}^{2} - \Omega_{1} R_{1}^{2}}{R_{2}^{2} - R_{1}^{2}},
    \qquad
    B = \frac{R_{1}^{2} R_{2}^{2} (\Omega_{1} - \Omega_{2})}{R_{2}^{2} - R_{1}^{2}}.
    \label{eq:classical_constants}
\end{equation}
The flux of angular momentum through a cylinder of radius $r$ is independent
of $r$ and is characterized by
\begin{equation}
    \Kcl \equiv -\eta r^{3}\Omcl' = 2\eta B,
    \label{eq:classical_flux}
\end{equation}
in terms of which the torque per unit length transmitted from one cylinder to
the other is $2\pi\Kcl=4\pi\eta B$ \cite{LandauLifshitz1987}.
\Cref{eq:classical_profile,eq:classical_constants,eq:classical_flux} serve as
the nonrelativistic reference for the relativistic formulations below;
accordingly, torques are reported as ratios $\mathcal{R}=K_{L}/\Kcl$.

\section{Relativistic geometry and kinematics}
\label{sec:kinematics}

\subsection{Cylindrical Minkowski geometry}
\label{subsec:geometry}

Since gravity is excluded, the background spacetime is Minkowski space, whose
line element in cylindrical coordinates is
\begin{equation}
    \dd s^{2} = -\dd t^{2} + \dd r^{2} + r^{2}\, \dd \phi^{2} + \dd z^{2},
    \label{eq:metric}
\end{equation}
so that $g_{\mu\nu}=\mathrm{diag}(-1,1,r^{2},1)$ and
$g^{\mu\nu}=\mathrm{diag}(-1,1,r^{-2},1)$. Since $g_{\phi\phi}=r^{2}$ is the
only coordinate-dependent component, the Christoffel symbols
$\Gamma^{\rho}_{\mu\nu}=\tfrac12 g^{\rho\tau}(\pd_{\mu}g_{\nu\tau}
+\pd_{\nu}g_{\mu\tau}-\pd_{\tau}g_{\mu\nu})$ have only two independent
nonvanishing components,
\begin{equation}
\begin{aligned}
    \Gamma^{r}_{\phi\phi} &= -\tfrac{1}{2}g^{rr}\pd_{r}g_{\phi\phi} = -r,\\
    \Gamma^{\phi}_{r\phi} = \Gamma^{\phi}_{\phi r}
    &= \tfrac{1}{2}g^{\phi\phi}\pd_{r}g_{\phi\phi} = \frac{1}{r}.
\end{aligned}
    \label{eq:christoffel}
\end{equation}

\subsection{Four-velocity}
\label{subsec:velocity}

The base flow is assumed to be stationary, axisymmetric, and independent of
$z$, $\pd_{t}=\pd_{\phi}=\pd_{z}=0$, with purely azimuthal motion,
$U^{r}=U^{z}=0$. Writing $\dd\phi/\dd t=\Omega(r)$ gives $U^{\phi}=\Omega U^{t}$,
and the normalization $U_{\mu}U^{\mu}=-1$ becomes
$-(U^{t})^{2}+r^{2}\Omega^{2}(U^{t})^{2}=-1$. Hence
\begin{equation}
    U^{\mu} = \bigl(\gamma,\, 0,\, \gamma\Omega,\, 0\bigr),
    \qquad
    \gamma(r) = \bigl(1 - r^{2}\Omega(r)^{2}\bigr)^{-1/2},
    \label{eq:four_velocity}
\end{equation}
with covariant components $U_{\mu}=(-\gamma,0,r^{2}\gamma\Omega,0)$. The
physical azimuthal velocity is $\vphi=r\Omega$, and $\gamma$ is real only if
$r|\Omega(r)|<1$ throughout the gap; in particular, both walls must be
subluminal. The wall speeds are denoted by $v_{i}=R_{i}\Omega_{i}$, with
$|v_{i}|<1$, and the corresponding Lorentz factors by
$\gamma_{i}=\gamma(R_{i})$. The kinematic
quantities below depend only on $U^{\mu}$ and $g_{\mu\nu}$, and they therefore
apply to both formulations of \cref{sec:conservation}.

\subsection{Expansion, four-acceleration, and shear}
\label{subsec:shear}

The expansion scalar
$\theta=\cd_{\mu}U^{\mu}=\pd_{\mu}U^{\mu}+\Gamma^{\mu}_{\mu\rho}U^{\rho}$
vanishes: the first term because $U^{\mu}$ depends only on $r$ and $U^{r}=0$,
the second because $\Gamma^{\mu}_{\mu\rho}$ is nonzero only for $\rho=r$.
Hence
\begin{equation}
    \theta = 0
    \label{eq:expansion}
\end{equation}
identically, without any incompressibility assumption. The four-acceleration
$a^{\mu}=U^{\rho}\cd_{\rho}U^{\mu}$ has a single nonzero component,
\begin{equation}
    a^{r} = \Gamma^{r}_{\phi\phi}\,(U^{\phi})^{2} = -r\gamma^{2}\Omega^{2},
    \label{eq:acceleration}
\end{equation}
which is the relativistic centripetal acceleration.

With the spatial projector $h_{\mu}{}^{\nu}=\delta_{\mu}{}^{\nu}+U_{\mu}U^{\nu}$
and $\theta=0$, the shear tensor is
\begin{equation}
    \sigma_{\mu\nu} = h_{\mu}{}^{\alpha} h_{\nu}{}^{\beta}\,
    \cd_{(\alpha}U_{\beta)},
    \qquad
    \cd_{(\alpha}U_{\beta)} = \tfrac12\bigl(\cd_{\alpha}U_{\beta}+\cd_{\beta}U_{\alpha}\bigr).
    \label{eq:shear_def}
\end{equation}
Its only nonzero components, derived in \cref{app:shear}, are
\begin{equation}
    \sigma^{r\phi} = \tfrac12\gamma^{3}\Omega'(r),
    \qquad
    \sigma^{rt} = r^{2}\Omega\,\sigma^{r\phi},
    \label{eq:shear_component}
\end{equation}
and the scalar needed for the viscous heating rate is
$\sigma_{\mu\nu}\sigma^{\mu\nu}=\tfrac12 r^{2}\gamma^{4}\Omega'^{2}$. The factor
$\gamma^{3}$ in $\sigma^{r\phi}$ arises from the projector component
$h_{\phi}{}^{\phi}=\gamma^{2}$ combined with the derivative of $\gamma\Omega$;
it is a purely kinematic effect and is common to both formulations.

\section{Conservation laws and radial heat conduction}
\label{sec:conservation}

The divergence identities used below are collected in \cref{app:identities}.
Throughout, the shear viscosity $\eta$ and, in \cref{subsec:heat_coupled}, the
thermal conductivity $\kappa$ are taken to be constant; the limits of this
assumption are discussed in \cref{subsec:thermal}.

\subsection{\texorpdfstring{Zero-heat-flux approximation ($q^{r}=0$)}{Zero-heat-flux approximation (q = 0)}}
\label{subsec:benchmark}

Without heat flux, the stress-energy tensor of a viscous fluid is
\begin{equation}
    T^{\alpha\beta} = e\,U^{\alpha}U^{\beta} + P h^{\alpha\beta}
    - 2\eta\,\sigma^{\alpha\beta} - \zeta\theta h^{\alpha\beta},
    \label{eq:Tmunu_zero}
\end{equation}
where $e$ is the energy density, $P$ the pressure, $\eta$ and $\zeta$ the
shear and bulk viscosities, and $h^{\alpha\beta}=g^{\alpha\beta}+U^{\alpha}U^{\beta}$.
The bulk term vanishes by \cref{eq:expansion}. Since $U^{r}=0$ and
$g^{r\phi}=g^{rt}=0$, the off-diagonal components with an index $r$ are purely
viscous,
\begin{equation}
    T^{r\phi} = -\eta\gamma^{3}\Omega',
    \qquad
    T^{rt} = r^{2}\Omega\, T^{r\phi},
    \label{eq:Trphi}
\end{equation}
while $T^{rr}=P$ and $T^{\phi\phi}=(e+P)\gamma^{2}\Omega^{2}+P/r^{2}$.

\paragraph{Angular momentum.}
The $\phi$ component of the conservation law reduces to
$\cd_{\alpha}T^{\alpha\phi}=r^{-3}\,\dd(r^{3}T^{r\phi})/\dd r=0$, so that
$r^{3}T^{r\phi}$ is constant. Denoting this constant by $K_{L}^{(0)}$,
\begin{equation}
    \frac{\dd}{\dd r}\bigl(\eta r^{3}\gamma^{3}\Omega'\bigr) = 0,
    \qquad
    \eta r^{3}\gamma^{3}\Omega' = -K_{L}^{(0)},
    \label{eq:zero_flux_ode}
\end{equation}
or, equivalently,
\begin{equation}
    \frac{\dd\Omega}{\dd r}
    = -\frac{K_{L}^{(0)}}{\eta r^{3}}\bigl(1-r^{2}\Omega^{2}\bigr)^{3/2}.
    \label{eq:zero_flux_explicit}
\end{equation}
Together with the no-slip conditions $\Omega(R_{1})=\Omega_{1}$ and
$\Omega(R_{2})=\Omega_{2}$, \cref{eq:zero_flux_explicit} is a nonlinear
boundary-value problem for $\Omega(r)$ in which $K_{L}^{(0)}$ is an
eigenvalue. Its solution is denoted by $\Omz(r)$, and
$\mathcal{R}_{0}=K_{L}^{(0)}/\Kcl$ is the corresponding torque ratio.
\Cref{eq:zero_flux_explicit} contains neither $e$ nor $P$, so that $\Omz(r)$
is independent of the density and of the equation of state; the inertia of the
fluid enters only the radial balance below and the stability of the flow.

\paragraph{Radial momentum.}
The $r$ component gives
$\cd_{\alpha}T^{\alpha r}=r^{-1}\,\dd(rT^{rr})/\dd r-rT^{\phi\phi}=0$, that is,
\begin{equation}
    \frac{\dd P}{\dd r} = (e+P)\, r\gamma^{2}\Omega^{2}.
    \label{eq:radial_balance}
\end{equation}
The pressure gradient balances the centripetal acceleration of the inertia
$e+P$. Once $\Omega(r)$ is known, \cref{eq:radial_balance} determines $P(r)$
for a given equation of state but does not feed back into the angular-momentum
equation.

\paragraph{Energy.}
The $t$ component is $\cd_{\alpha}T^{\alpha t}=r^{-1}\,\dd(rT^{rt})/\dd r$.
Using \cref{eq:Trphi} and then \cref{eq:zero_flux_ode},
\begin{equation}
    \cd_{\alpha}T^{\alpha t}
    = \frac{\Omega}{r}\frac{\dd}{\dd r}\bigl(r^{3}T^{r\phi}\bigr)
      + r^{2}\Omega' T^{r\phi}
    = -\eta r^{2}\gamma^{3}\Omega'^{2}
    = -\frac{\Phi}{\gamma},
    \label{eq:energy_violation}
\end{equation}
where $\Phi=2\eta\,\sigma_{\mu\nu}\sigma^{\mu\nu}=\eta r^{2}\gamma^{4}\Omega'^{2}$
is the rate of viscous heating in the fluid rest frame. The energy equation is
therefore satisfied only where $\Omega'=0$, that is, only for rigid rotation.
For any sheared profile, \cref{eq:energy_violation} identifies the quantity
that the truncated model omits: the heat generated by viscosity at the rate
$\Phi$ (per unit proper volume and proper time) is neither stored nor
transported. Consequently, the zero-heat-flux solution can be stationary only
if each fluid element is in contact with an external sink that removes this
energy locally. This is the cylindrical counterpart of the inconsistency
identified in the planar case by Gavassino \emph{et al.}\
\cite{Gavassino2026}, and it motivates the heat-conducting formulation
below.

\subsection{Heat-conducting model in the Eckart frame}
\label{subsec:heat_coupled}

The Eckart frame \cite{Eckart1940} is defined by the particle current,
$N^{\mu}=nU^{\mu}$, where $n$ is the rest-frame particle density. Its
conservation, $\cd_{\mu}N^{\mu}=n\theta+U^{\mu}\pd_{\mu}n=0$, is satisfied
identically by the present flow, since $\theta=0$ and $n=n(r)$ is constant
along the azimuthal streamlines. The model therefore describes a fluid with a
conserved particle number; for a fluid without a conserved charge, the Eckart
frame is not defined and the formulation must be modified
\cite{GavassinoSchlichtingDenicol2026}.

A heat flux $q^{\mu}$ orthogonal to $U^{\mu}$ is added to
\cref{eq:Tmunu_zero},
\begin{equation}
    T^{\alpha\beta} = e\,U^{\alpha}U^{\beta} + P h^{\alpha\beta}
    - 2\eta\,\sigma^{\alpha\beta} + q^{\alpha}U^{\beta} + U^{\alpha}q^{\beta}.
    \label{eq:Tmunu_eckart}
\end{equation}
For the present symmetry the heat flux is radial, $q^{\mu}=(0,q(r),0,0)$, and
satisfies $U_{\mu}q^{\mu}=0$ automatically. The heat flux is related to the temperature
$T(r)$ by Eckart's law,
\begin{equation}
    q^{\mu} = -\kappa\, h^{\mu\nu}\bigl(\pd_{\nu}T + T a_{\nu}\bigr),
    \label{eq:eckart_law}
\end{equation}
whose radial component, by \cref{eq:acceleration} and $h^{rr}=1$, is
\begin{equation}
    q = -\kappa\bigl(T' - r\gamma^{2}\Omega^{2}\, T\bigr).
    \label{eq:eckart_radial}
\end{equation}
The second term is the inertial (Tolman) contribution. In a static
gravitational field it yields the Tolman--Ehrenfest relation
\cite{TolmanEhrenfest1930}, and by the equivalence principle the same term
accounts for the centripetal acceleration of a rotating fluid. In a rotating
fluid, therefore, a radial temperature gradient alone does not determine the
heat flux, since in rigidly rotating equilibrium the temperature is not
uniform but proportional to the local Lorentz factor
\cite{GavassinoAntonelli2025}.

\paragraph{First integrals.}
With the heat flux, the components of \cref{eq:Tmunu_eckart} with an index $r$
become
\begin{equation}
    T^{r\phi} = -\eta\gamma^{3}\Omega' + \gamma\Omega\, q,
    \qquad
    T^{rt} = -\eta r^{2}\gamma^{3}\Omega\Omega' + \gamma q,
    \label{eq:T_heat}
\end{equation}
while $T^{rr}=P$ and $T^{\phi\phi}$ are unchanged, because $q^{\mu}$ has no $t$
or $\phi$ component and $U^{r}=0$. By \cref{eq:div_t,eq:div_phi}, the equations
$\cd_{\alpha}T^{\alpha t}=0$ and $\cd_{\alpha}T^{\alpha\phi}=0$ integrate to
\begin{equation}
    r\,T^{rt} = K_{E},
    \qquad
    r^{3}\,T^{r\phi} = K_{L},
    \label{eq:first_integrals}
\end{equation}
where $K_{E}$ and $K_{L}$ are constants: $2\pi K_{E}$ is the energy and
$2\pi K_{L}$ the angular momentum transported outwards through a cylinder of
radius $r$, per unit length and unit time. In contrast to
\cref{subsec:benchmark}, both conservation laws can now be satisfied
simultaneously, since $q(r)$ provides the additional unknown function.

\paragraph{Reduced angular-momentum equation.}
The first relation in \cref{eq:first_integrals} gives
$\gamma q=K_{E}/r+\eta r^{2}\gamma^{3}\Omega\Omega'$. Substituting this into the
second and using $1-r^{2}\Omega^{2}=\gamma^{-2}$ yields
\begin{equation}
    \eta r^{3}\gamma\,\Omega' = r^{2}\Omega K_{E} - K_{L},
    \label{eq:reduced}
\end{equation}
and, after inserting \cref{eq:reduced} back into the expression for $q$,
\begin{equation}
    q = \frac{\gamma}{r}\bigl(K_{E} - \Omega K_{L}\bigr).
    \label{eq:q_integrals}
\end{equation}
Both relations are exact. A comparison of \cref{eq:reduced} with
\cref{eq:zero_flux_ode} shows two effects of the heat flux. The term
$\gamma\Omega q$ in $T^{r\phi}$, which describes the angular momentum carried
by heat, cancels a factor $\gamma^{-2}$ of the viscous term, so that $\gamma^{3}$
is replaced by $\gamma$; in addition, the energy flux enters through the term
$r^{2}\Omega K_{E}$. For numerical integration it is convenient to use the
rapidity $y$, defined by $r\Omega=\tanh y$, so that $\gamma=\cosh y$ and
\cref{eq:reduced} becomes
\begin{equation}
    \frac{\dd y}{\dd r} = \frac{\sinh y\cosh y}{r}
    + \frac{\cosh y}{\eta r^{2}}\bigl(rK_{E}\tanh y - K_{L}\bigr).
    \label{eq:rapidity}
\end{equation}

\paragraph{Temperature equation.}
Equating \cref{eq:q_integrals} with \cref{eq:eckart_radial} gives
\begin{equation}
    \kappa\bigl(T' - r\gamma^{2}\Omega^{2}\,T\bigr)
    = -\frac{\gamma}{r}\bigl(K_{E} - \Omega K_{L}\bigr).
    \label{eq:temperature}
\end{equation}
Since $\gamma'/\gamma=r\gamma^{2}\Omega^{2}+r^{2}\gamma^{2}\Omega\Omega'$, the
left-hand side can be written as
$\kappa[\gamma(T/\gamma)'+r^{2}\gamma^{2}\Omega\Omega'\,T]$. For rigid
rotation the heat flux therefore vanishes when $T/\gamma$ is uniform, which is
the rotating form of the Tolman--Ehrenfest relation, whereas in a sheared flow
the second term drives a heat flux even when $T/\gamma$ is uniform.

\paragraph{Radial balance and equation of state.}
Since $T^{rr}$ and $T^{\phi\phi}$ are unchanged, the radial component again
gives \cref{eq:radial_balance}. For constant $\eta$ and $\kappa$, moreover,
neither \cref{eq:reduced} nor \cref{eq:temperature} contains $e$ or $P$, so
that the profiles $\Omq(r)$ and $T(r)$ are independent of the equation of state, which
enters only through the pressure and density distributions that follow from
\cref{eq:radial_balance}.

\paragraph{Boundary conditions.}
\Cref{eq:reduced,eq:temperature} are first-order equations for $\Omega(r)$ and
$T(r)$ containing the two unknown constants $K_{E}$ and $K_{L}$. The four
unknowns are fixed by the no-slip conditions $\Omega(R_{1})=\Omega_{1}$ and
$\Omega(R_{2})=\Omega_{2}$ together with the two wall temperatures, discussed
in \cref{subsec:thermal}.

\paragraph{Energy budget.}
\Cref{eq:q_integrals} can be written as $K_{E}=\Omega K_{L}+rq/\gamma$, which
holds at every radius. Evaluating it at both walls and subtracting gives
\begin{equation}
    2\pi(\Omega_{1}-\Omega_{2})K_{L}
    = 2\pi\left[\frac{R_{2}\,q(R_{2})}{\gamma(R_{2})}
    - \frac{R_{1}\,q(R_{1})}{\gamma(R_{1})}\right].
    \label{eq:budget}
\end{equation}
The left-hand side is the net mechanical power, per unit length, supplied by
the cylinders, whereas the right-hand side is the net heat leaving the fluid
through the two walls. The factor $1/\gamma$ has a direct interpretation: at
wall $i$ the four-velocity is $U^{\mu}=\gamma(R_{i})\,\xi_{i}^{\mu}$, where
$\xi_{i}=\pd_{t}+\Omega_{i}\pd_{\phi}$ is a Killing vector of Minkowski space,
and $R_{i}q(R_{i})/\gamma(R_{i})=K_{E}-\Omega_{i}K_{L}$ is the flux of the
conserved energy associated with $\xi_{i}$, that is, the heat current measured
in the frame that co-rotates with that wall. In a stationary state,
therefore, all the work done on the fluid is removed as heat through the
boundaries, which is precisely the balance that the zero-heat-flux model
cannot satisfy. In the Landau frame, whose velocity follows the energy flux,
the fluid velocity differs from $U^{\mu}$ by $q^{\mu}/(e+P)$ to first order in
the dissipative fluxes \cite{Gavassino2026}. The Landau velocity therefore has
a radial component $q/(e+P)$, and the heat removed through the walls appears in
that frame as a flow of energy across them. This is the cylindrical
counterpart of the picture emphasized in Ref.~\cite{Gavassino2026}, in which
the fluid removes the energy generated by viscous heating by streaming across
the boundaries.

\paragraph{Entropy production.}
With the entropy current $S^{\mu}=sU^{\mu}+q^{\mu}/T$, the first-order theory
gives
\begin{equation}
    \cd_{\mu}S^{\mu} = \frac{\Phi}{T} + \frac{q^{2}}{\kappa T^{2}} \geq 0,
    \label{eq:entropy}
\end{equation}
where $\Phi$ is the viscous heating rate of \cref{eq:energy_violation} and
$q^{2}=q_{\mu}q^{\mu}$ \cite{Eckart1940,RezzollaZanotti2013}. Both terms are
nonnegative for $\eta,\kappa>0$ and $T>0$.

The solution of the heat-conducting problem is denoted by $\Omq(r)$, and its
torque ratio by $\mathcal{R}_{q}=K_{L}/\Kcl$.

\subsection{Thermal boundary conditions and the energy flux}
\label{subsec:thermal}

The walls are assumed to impose no velocity slip and no temperature jump. The
relative importance of conduction is characterized by the dimensionless
parameter
\begin{equation}
    \lambda = \frac{\kappa T_{w}}{\eta c^{2}},
    \label{eq:lambda}
\end{equation}
where $T_{w}$ is a reference wall temperature. For ordinary liquids $\lambda$
is extremely small ($\lambda\approx2\times10^{-12}$ for water at room
temperature), whereas in a relativistically hot gas both $\eta c^{2}$ and
$\kappa T$ are set by the enthalpy density and the same mean free path \cite{CercignaniKremer2002}, so that $\lambda$ is of order unity. In what follows
it is convenient to use $\Theta\equiv\kappa T/(\eta c^{2})$, which equals
$\lambda T/T_{w}$.

\paragraph{Energy flux.}
Since \cref{eq:temperature} is linear in $T$, it can be integrated with the
integrating factor
\begin{equation}
    \mu(r) = \exp\!\Bigl(-\int_{R_{1}}^{r} r'\gamma^{2}\Omega^{2}\,\dd r'\Bigr),
    \label{eq:mu}
\end{equation}
with which it becomes $\kappa\,(\mu T)'=-\mu\gamma(K_{E}-\Omega K_{L})/r$.
Integration between the walls then gives the exact relation
\begin{equation}
\begin{split}
    K_{E}\int_{R_{1}}^{R_{2}}\mu\gamma\,\frac{\dd r}{r}
    &= K_{L}\int_{R_{1}}^{R_{2}}\mu\gamma\Omega\,\frac{\dd r}{r}
     + \kappa\,\delta,\\
    \delta &\equiv T(R_{1}) - \mu(R_{2})\,T(R_{2}).
\end{split}
    \label{eq:KE_exact}
\end{equation}
This relation connects the energy flux with the angular-momentum flux, the
profile, and the wall temperatures; in particular, the wall temperatures enter
only through the mismatch $\delta$.

\paragraph{Conductivity-independent solutions.}
The conductivity term of \cref{eq:KE_exact} vanishes for every $\kappa$ if the
wall temperatures satisfy
\begin{equation}
    \frac{T(R_{2})}{T(R_{1})} = \frac{1}{\mu(R_{2})}
    = \exp\!\Bigl(\int_{R_{1}}^{R_{2}} r\gamma^{2}\Omega^{2}\,\dd r\Bigr).
    \label{eq:balanced}
\end{equation}
Walls with this property are referred to below as balanced. In this case
$K_{E}=K_{L}\langle\Omega\rangle_{\mu}$, where $\langle\Omega\rangle_{\mu}$ is
the average of $\Omega$ with weight $\mu\gamma/r$, and since \cref{eq:reduced}
does not contain $\kappa$ either, $\Omq(r)$, $K_{L}$, and $K_{E}$ are
independent of $\kappa$. Integrating \cref{eq:temperature} once more then
gives the temperature,
\begin{equation}
    T(r) = \frac{T(R_{1})}{\mu(r)} + \frac{\eta c^{2}}{\kappa}\,\Theta_{0}(r),
    \label{eq:T_balanced}
\end{equation}
where $\Theta_{0}$ is the solution of \cref{eq:temperature}, written for
$\Theta$, that vanishes at both walls. The first term describes the profile
for which the bracket in \cref{eq:eckart_radial} vanishes and which therefore
carries no heat, whereas the second describes the temperature rise produced by
viscous heating, which decreases as $1/\lambda$.

The balance condition has two familiar limiting forms. For rigid rotation,
$(\ln\gamma)'=r\gamma^{2}\Omega^{2}$, so that $1/\mu=\gamma/\gamma(R_{1})$ and
\cref{eq:balanced} becomes the Tolman--Ehrenfest relation $T\propto\gamma$
\cite{TolmanEhrenfest1930,GavassinoAntonelli2025}. For planar flow, $\mu=1$,
and \cref{eq:balanced} reduces to equal plate temperatures, for which the
planar profile is likewise independent of $\kappa$ \cite{Gavassino2026}.
Balanced walls therefore play in the cylindrical problem the role that equal
plate temperatures play in the planar one. Since $\mu$ depends on the profile,
however, \cref{eq:balanced} is a condition on the solution rather than an
externally prescribed boundary condition; in practice, $\mu(R_{2})$ follows
from the $\kappa$-independent profile itself. For $R_{2}/R_{1}=2$ and
$\Omega_{2}=0$, the required ratio $T(R_{2})/T(R_{1})$ is $1.07$, $2.16$, and
$27.2$ at $v_{1}=0.5c$, $0.9c$, and $0.99c$, respectively.

The same solution is obtained in the limit $\lambda\to0$ at fixed wall
temperatures, since the term $\kappa\delta$ in \cref{eq:KE_exact} then vanishes
as well.
Although $T-T_{w}$ grows as $1/\lambda$ in this limit, the product $\kappa T$
remains finite, and the inertial term $\kappa r\gamma^{2}\Omega^{2}T$ of
\cref{eq:eckart_radial} remains comparable to $\kappa T'$; consequently, it
survives through the weight $\mu$. Neglecting it altogether ($\mu=1$), as the
planar analogy might suggest, is therefore justified only when
$\int r\gamma^{2}\Omega^{2}\,\dd r\ll1$, that is, at small wall speeds or in a
narrow gap. More specifically, for $R_{2}/R_{1}=2$ and $\Omega_{2}=0$, setting
$\mu=1$ changes the torque ratio by about $10^{-4}$ at $v_{1}=0.3c$ and by
$0.1\%$ at $0.5c$, whereas it gives $2.44$ instead of $2.71$ at $0.9c$ and
underestimates the torque ratio by factors of $2.3$ and $7.2$ at $0.99c$ and
$0.999c$.

\paragraph{Other wall temperatures.}
Balanced walls are compared below with two simpler protocols: equal wall
temperatures, $T(R_{1})=T(R_{2})=T_{w}$, and equal values of $T/\gamma$ at the
walls, $T(R_{i})=\gamma(R_{i})\,T_{w}$. Although the second protocol is
motivated by the Tolman--Ehrenfest relation of rigid rotation, it does not
describe an equilibrium when $\Omega_{1}\neq\Omega_{2}$; in particular, for
$\Omega_{2}=0$ it makes the inner wall hotter, whereas balanced walls require a
hotter outer wall. All three protocols are idealizations, since real walls
need not act as ideal thermostats \cite{GavassinoSchlichtingDenicol2026}. For
both simpler protocols $\delta\neq0$ in general, and \cref{eq:KE_exact} adds
to $K_{E}$ a contribution proportional to $\lambda$. Its effect on the torque follows from
integrating \cref{eq:reduced} across the gap,
\begin{equation}
    K_{L}\int_{R_{1}}^{R_{2}}\frac{\dd r}{\gamma r^{3}}
    = K_{E}\int_{R_{1}}^{R_{2}}\frac{\Omega\,\dd r}{\gamma r}
    + \eta(\Omega_{1}-\Omega_{2}),
    \label{eq:KL_KE}
\end{equation}
which shows that, at given wall speeds, a change in the energy flux changes the
angular-momentum flux in the same direction when
$\int_{R_{1}}^{R_{2}}\Omega\,\dd r/(\gamma r)>0$ and in the opposite direction
when this integral is negative. It follows that, to first order in $\delta$,
the torque ratio changes with $\lambda$ with the sign of
$\delta\,\Kcl\int_{R_{1}}^{R_{2}}\Omega\,\dd r/(\gamma r)$. When $\Omega\geq0$
throughout the gap, as in all cases of \cref{tab:heat}, the integral is
positive, and the torque ratio increases with $\lambda$ when $\delta$ and
$\Kcl$ have the same sign. With $\Omega_{2}=0$, $\delta$ is positive for both
protocols, $\delta=T_{w}[1-\mu(R_{2})]$ for equal wall temperatures and
$\delta=T_{w}[\gamma_{1}-\mu(R_{2})]$, which is larger, for equal
$T/\gamma$, so that the torque ratio increases with $\lambda$. In
counter-rotating flows, $\Omega_{1}\Omega_{2}<0$, the integral can vanish or
become negative. For $(\Omega_{1},\Omega_{2})=(0.3,-0.3)$, where it is
negative, the torque ratio at $\lambda=1$ decreases from $1.047$ to $1.037$
with equal wall temperatures ($\delta>0$) and increases to $1.062$ with equal
$T/\gamma$ ($\delta<0$), in agreement with this form of the rule. Since the
profile itself also changes with $\lambda$, the rule is not exact; it has been
confirmed numerically for $\lambda=1$ in all eight combinations of case and
protocol in \cref{tab:heat}
and in the counter-rotating cases $(0.3,-0.3)$ and $(0.5,-0.2)$, and, at
$v_{1}=0.9c$, for $\lambda$ up to $10$ (\cref{subsec:profiles}).

\paragraph{Temperature variation and the Brinkman number.}
The assumption of constant transport coefficients requires the temperature to
vary little across the gap. A first estimate follows from balancing the
viscous heating rate, $\sim\eta v^{2}/d^{2}$ for a wall speed $v$ and gap
width $d=R_{2}-R_{1}$, against conduction, $\sim\kappa\,\Delta T/d^{2}$, which
gives $\Delta T/T_{w}\sim\Br$, with the Brinkman number
\begin{equation}
    \Br = \frac{\eta v_{w}^{2}}{\kappa T_{w}} = \frac{v_{w}^{2}}{\lambda c^{2}},
    \label{eq:brinkman}
\end{equation}
where $v_{w}$ is the larger wall speed. At wall speeds comparable to $c$,
$\Br\lesssim1$ therefore requires $\lambda\gtrsim v_{w}^{2}/c^{2}$, whereas a
fluid with $\lambda\ll1$ is heated far above the wall temperature; in
particular, the limit $\lambda\to0$ at fixed wall temperatures corresponds to
$\Br\to\infty$. At large Lorentz factors, however, two further effects
appear (\cref{subsec:profiles}). First, the estimate underestimates the
heating. For $\lambda=1$ and equal wall temperatures, the maximum rise above
the wall temperature, $\Delta T_{\max}$, is $0.19\,\Br\,T_{w}$ and
$0.40\,\Br\,T_{w}$ in the cases of \cref{tab:heat} with $v_{w}=0.5c$, but
$1.5\,\Br\,T_{w}$ at $v_{1}=0.9c$ and $17\,\Br\,T_{w}$ at $0.99c$. Second, the
inertial term imposes a temperature variation of its own. With equal wall
temperatures, $\Delta T_{\max}$ decreases only slowly with $\lambda$ and is
still $0.39\,T_{w}$ at $v_{1}=0.9c$ and $3.3\,T_{w}$ at $0.99c$ for
$\lambda=30$, whereas with balanced walls the viscous rise decreases as
$1/\lambda$ but the wall temperatures themselves differ by the factor
$1/\mu(R_{2})$. Consequently, for the protocols considered here, the
temperature varies across the gap by about $40\%$ or more at $v_{1}=0.9c$ and
by a factor of several at $0.99c$, for all values of $\lambda$ considered
(up to $30$). At
these speeds the results should be
interpreted as properties of the constant-coefficient model rather than as
quantitative predictions for a real fluid, whose transport coefficients depend
on temperature.

\subsection{Nonrelativistic, weakly relativistic, and narrow-gap limits}
\label{subsec:limits}

\paragraph{Zero-heat-flux model.}
Restoring $c$, $\gamma=(1-r^{2}\Omega^{2}/c^{2})^{-1/2}$. For
$r|\Omega|/c\ll1$, $\gamma\to1$ and \cref{eq:zero_flux_ode} reduces to
$\dd(\eta r^{3}\Omega')/\dd r=0$, whose solution for constant $\eta$ is the
classical profile of \cref{eq:classical_profile}, with $K_{L}^{(0)}\to\Kcl$.
To the next order, $\gamma^{3}\simeq1+\tfrac32 r^{2}\Omega^{2}/c^{2}$. Writing
$\Omega=\Omcl+O(c^{-2})$ in the correction term and integrating
\cref{eq:zero_flux_ode} across the gap with the boundary values held fixed
gives
\begin{equation}
    \mathcal{R}_{0} = \frac{K_{L}^{(0)}}{\Kcl}
    = 1 + \frac{3}{2c^{2}}\,\frac{I_{2}}{I_{3}} + O(c^{-4}),
    \label{eq:pn_zero}
\end{equation}
with
\begin{equation}
    I_{2} = \int_{R_{1}}^{R_{2}} \Omcl^{2}\,\frac{\dd r}{r},
    \qquad
    I_{3} = \int_{R_{1}}^{R_{2}} \frac{\dd r}{r^{3}}
          = \frac{R_{2}^{2}-R_{1}^{2}}{2R_{1}^{2}R_{2}^{2}} .
    \label{eq:pn_integrals}
\end{equation}
For $R_{1}=1$, $R_{2}=2$, $\Omega_{1}=0.5$, $\Omega_{2}=0$ (and $c=1$),
$I_{2}=0.04009$ and $I_{3}=0.375$, so that the leading correction is
$\tfrac32 I_{2}/I_{3}=0.160$; equivalently,
$(\mathcal{R}_{0}-1)c^{2}/v_{1}^{2}\to0.6414$ as $v_{1}\to0$ in this geometry.

\paragraph{Heat-conducting model.}
By \cref{eq:KE_exact}, $K_{E}$ is of order $K_{L}\Omega$, apart from the term
$\kappa\delta$, so that the energy-flux term $r^{2}\Omega K_{E}$ in
\cref{eq:reduced} is of order $(v/c)^{2}$ relative to $K_{L}$. In the
nonrelativistic limit $\gamma\to1$, \cref{eq:reduced} reduces to
$\eta r^{3}\Omega'=-K_{L}$, and the classical profile is recovered with
$K_{L}\to\Kcl$. To the next order, $\gamma\simeq1+\tfrac12 r^{2}\Omega^{2}/c^{2}$.
Integrating \cref{eq:reduced} across the gap with $\Omega=\Omcl$ in the
correction terms gives
\begin{equation}
    \mathcal{R}_{q} = 1 + \frac{1}{c^{2}I_{3}}
    \left[\frac{\bar{K}_{E}}{\Kcl}\,I_{1} + \frac12 I_{2}\right] + O(c^{-4}),
    \label{eq:pn_heat}
\end{equation}
with $I_{1}=\int_{R_{1}}^{R_{2}}\Omcl\,\dd r/r$,
and $\bar{K}_{E}$ the leading-order energy flux. Setting $\gamma=1$ and
$\mu\simeq1-\int r\Omcl^{2}\,\dd r$ in \cref{eq:KE_exact},
\begin{equation}
    \bar{K}_{E} = \frac{\Kcl I_{1}}{\ell}
    + \frac{\kappa}{\ell}\Bigl[T(R_{1}) - T(R_{2}) + T(R_{2})\,I_{T}\Bigr],
    \label{eq:KE0}
\end{equation}
with $I_{T}=\int_{R_{1}}^{R_{2}} r\,\Omcl^{2}\,\dd r$ and
$\ell=\ln(R_{2}/R_{1})$. To leading order, the bracket is the mismatch
$\delta$. It vanishes for balanced walls, for which
$T(R_{2})=T(R_{1})(1+I_{T})$ at this order, whereas it equals $T_{w}I_{T}$ for
equal wall temperatures and $T_{w}(I_{T}+\tfrac12R_{1}^{2}\Omega_{1}^{2}/c^{2})$
for equal $T/\gamma$ with $\Omega_{2}=0$.
For the test case of \cref{eq:pn_integrals}, $I_{1}=0.13448$,
$I_{T}=0.05430$, and $\ell=0.69315$, so that $\mathcal{R}_{q}-1=0.1230$ for the
$\kappa$-independent solution, compared with $0.160$ for the zero-heat-flux
model; the conductivity term adds $0.0211\,\lambda$ for equal wall
temperatures and $0.0696\,\lambda$ for equal $T/\gamma$. Per unit
$v_{1}^{2}/c^{2}$, the coefficients are $0.4921$ for the $\kappa$-independent
solution and, for $\lambda=1$, $0.5764$ with equal wall temperatures and
$0.7704$ with equal $T/\gamma$. At this order the correction of the
zero-heat-flux model ($0.6414$) therefore exceeds that of the
$\kappa$-independent solution by about $30\%$, whereas the correction of the
heat-conducting model for $\lambda>0$ depends on the wall temperatures and can
be the larger one.
These coefficients are compared with the numerical solutions in
\cref{subsec:checks}.

\paragraph{Narrow-gap limit.}
Let $r=R_{1}+x$ with $d=R_{2}-R_{1}\ll R_{1}$, and write $v=r\Omega$. Then
$r^{3}\Omega'=r^{2}v'-rv=R_{1}^{2}v'\,[1+O(d/R_{1})]$, because $v'\sim v/d$.
To leading order in $d/R_{1}$, \cref{eq:zero_flux_ode} becomes
$\eta\gamma^{3}v'=-K_{L}^{(0)}/R_{1}^{2}$, the planar equation of Rogava
\cite{Rogava1996}, for which the proper velocity $\gamma v$ is linear in $x$.
Similarly, \cref{eq:reduced} becomes
\begin{equation}
    \eta\gamma v' = v\,\mathcal{A} - \mathcal{C},
    \qquad
    \mathcal{A}=\frac{K_{E}}{R_{1}},\quad \mathcal{C}=\frac{K_{L}}{R_{1}^{2}},
    \label{eq:planar_reduced}
\end{equation}
and, since the inertial term in \cref{eq:temperature} is smaller than
$\kappa T'$ by a factor of order $d/R_{1}$ for fixed $\lambda$ and wall speeds,
the temperature equation becomes $\kappa T'=-\gamma(\mathcal{A}-v\mathcal{C})$.
These are the planar Eckart equations of Gavassino \emph{et al.}\
\cite{Gavassino2026}, with $\mathcal{A}$ and $\mathcal{C}$ the planar energy and
momentum fluxes; accordingly, both models reduce to their planar
counterparts.

For equal plate temperatures, the planar Eckart problem has a closed-form
solution. In the frame in which the plates move with velocities $\pm v_{p}$,
the solution is antisymmetric about the midplane, which requires
$\mathcal{A}=0$. \Cref{eq:planar_reduced} then gives $\eta\gamma v'=-\mathcal{C}$,
and with $v=\sin\vartheta$, so that $\gamma v'=\vartheta'$, the angle
$\vartheta$ is linear in $x$:
\begin{equation}
    v_{s}(x) = \sin\!\Bigl[\Bigl(1-\frac{2x}{d}\Bigr)\arcsin v_{p}\Bigr],
    \qquad
    \mathcal{C} = \frac{2\eta}{d}\arcsin v_{p},
    \label{eq:planar_symmetric}
\end{equation}
independently of $\kappa$. Since $\gamma v=\tan\vartheta$, this is the
solution given in Ref.~\cite{Gavassino2026} for the proper velocity. For
$\Omega_{2}=0$ the cylindrical problem corresponds to the frame in which the
outer plate is at rest, which is reached by the boost
$v=(v_{s}+v_{p})/(1+v_{s}v_{p})$ with
$v_{p}=\tanh\bigl(\tfrac12\operatorname{artanh}v_{1}\bigr)$. For the
zero-heat-flux model, $\gamma v$ decreases linearly from $\gamma_{1}v_{1}$ at
the inner wall to zero at the outer wall.

\section{Numerical method and results}
\label{sec:numerics}

\subsection{Dimensionless problem}
\label{subsec:dimensionless}

Lengths are measured in units of $R_{1}$, and units are chosen such that
$c=1$ and $\eta=1$. The zero-heat-flux problem is then specified by three
numbers, the radius ratio $R_{2}/R_{1}$ and the wall speeds
$\Omega_{1}R_{1}$ and $\Omega_{2}R_{2}$. The heat-conducting problem requires
in addition the parameter $\lambda$ of \cref{eq:lambda}, with the temperature
measured in units of $T_{w}$, and one of the three thermal protocols of
\cref{subsec:thermal}. Except in \cref{subsec:checks}, $R_{2}/R_{1}=2$.

\subsection{Solution method}
\label{subsec:method}

All numerical results were obtained with the code of Ref.~\cite{GelekvaCode2026},
which also reproduces every figure and table of this paper. The
zero-heat-flux problem was solved by shooting. Since integration from the
outer wall was found to be better conditioned at large wall speeds than
integration from the inner wall, \cref{eq:zero_flux_explicit} was integrated
inwards
from $\Omega(R_{2})=\Omega_{2}$ with an eighth-order Dormand--Prince method
\cite{HairerNorsettWanner1993} (relative tolerance $10^{-12}$), and
$K_{L}^{(0)}$ was determined by Brent's method \cite{Brent1973} so that
$\Omega(R_{1})=\Omega_{1}$.

For the heat-conducting problem, \cref{eq:rapidity,eq:temperature} were
written for the rapidity $y(r)$ and for $\Theta(r)$, together with
$M(r)=\int_{R_{1}}^{r}r'\gamma^{2}\Omega^{2}\,\dd r'$, so that $\mu=e^{-M}$. In
these variables the limit $\lambda\to0$ is regular, with
$\Theta(R_{1})=\Theta(R_{2})=0$, and balanced walls are imposed
self-consistently through $\Theta(R_{2})=\Theta(R_{1})\,e^{M(R_{2})}$. With
$y(R_{i})=\operatorname{artanh}(R_{i}\Omega_{i})$, $M(R_{1})=0$, and the two
thermal conditions, the problem was solved by fourth-order collocation with
adaptive mesh refinement \cite{KierzenkaShampine2001}, as implemented in
SciPy's \texttt{solve\_bvp} \cite{Virtanen2020} (tolerance $10^{-9}$),
with $K_{L}$ and $K_{E}$ as unknown parameters. Large wall speeds were reached
by continuation, in which both wall rapidities were increased in equal steps
from $2\%$ of their final values, starting from the classical profile; a
single solution at $v_{1}=0.999c$ required $268$ steps, and the torque curves
of \cref{subsec:profiles} were computed with $364$. For equal $T/\gamma$,
collocation with tolerance $10^{-9}$ exceeds the mesh limit above
$v_{1}\approx0.998c$, independently of the number of continuation steps; this
branch was therefore computed with tolerance $10^{-8}$, for which the change
from $10^{-7}$ is $3\times10^{-10}$ at $0.999c$. The same solutions were
obtained independently by continuation in the wall temperature, starting from
the equal-temperature solution at the same speed, and the two procedures agree
to $10^{-11}$ at $0.995c$, $0.998c$, and $0.999c$. Since, in addition, the
local slope of $\mathcal{R}_{q}$ against $\gamma_{1}$ varies smoothly up to
$0.999c$, the failure at the tighter tolerance is attributed to the
collocation mesh rather than to a turning point or to the loss of a
solution.

The accuracy of the solutions is examined in \cref{app:numerics} and
summarized in \cref{tab:checks}. In particular, the first integrals
\cref{eq:first_integrals} hold to $1.4\times10^{-7}$ between the collocation
nodes, the energy budget \cref{eq:budget} to machine precision, and the local
energy balance
$\cd_{\mu}q^{\mu}+q^{\mu}a_{\mu}=\Phi$, which follows from
$U_{\nu}\cd_{\mu}T^{\mu\nu}=0$, to $2.5\times10^{-12}$. Two-parameter shooting
reproduces $(K_{L},K_{E})$ to $5.4\times10^{-11}$, and balanced walls reproduce
the $\lambda\to0$ solution to $7\times10^{-12}$ for $\lambda$ up to $20$ (up
to $5$ at $v_{1}=0.99c$).

\subsection{Cases and stability of the base flow}
\label{subsec:cases}

The four cases in \cref{tab:zero,tab:heat} combine rotation of either
cylinder with moderate and large wall speeds, and their stability differs. By
Rayleigh's criterion, the classical inviscid flow is centrifugally stable if
$(r^{2}\Omega)^{2}$ does not decrease outwards \cite{LandauLifshitz1987}. For
$(\Omega_{1},\Omega_{2})=(0.5,0.2)$ and $(0,0.2)$, $r^{2}\Omega$ increases
from the inner to the outer wall and the flow is stable, whereas for
$\Omega_{2}=0$, $r^{2}\Omega$ falls to zero at the outer wall and the
criterion is violated. In the latter cases, the laminar solutions computed here
exist for all wall speeds considered but are realized physically only below a
critical Taylor number, above which Taylor vortices form \cite{Taylor1923}.
Since neither profile depends on the density or on the equation of state
(\cref{subsec:benchmark,subsec:heat_coupled}), the Reynolds number does not
affect the solutions; it enters through this stability question, which, like
the relativistic generalization of the criterion, is outside the scope of the
present work, and through the validity of the first-order description
(\cref{subsec:limitations}).

\begin{table}[t]
    \caption{\label{tab:zero}
    Zero-heat-flux model compared with the classical profile for
    $R_{2}/R_{1}=2$ and $c=\eta=1$: $\Delta\Omega_{0}\equiv\max_{r}|\Omz-\Omcl|$
    and $\mathcal{R}_{0}=K_{L}^{(0)}/\Kcl$. The second column states whether
    the classical Rayleigh criterion is satisfied (\cref{subsec:cases}).}
    \begin{ruledtabular}
    \begin{tabular}{lccc}
        $(\Omega_{1},\Omega_{2})$ & Rayleigh-stable & $\Delta\Omega_{0}$
        & $\mathcal{R}_{0}$ \\
        \hline
        $(0.50,\,0.20)$ & yes & $6.7\times10^{-3}$ & $1.4064$ \\
        $(0.00,\,0.20)$ & yes & $5.1\times10^{-3}$ & $1.0545$ \\
        $(0.50,\,0.00)$ & no  & $0.0294$ & $1.2122$ \\
        $(0.90,\,0.00)$ & no  & $0.318$  & $4.8905$ \\
    \end{tabular}
    \end{ruledtabular}
\end{table}

\begin{table}[t]
    \caption{\label{tab:heat}
    Torque ratio $\mathcal{R}_{q}=K_{L}/\Kcl$ of the heat-conducting model for
    $R_{2}/R_{1}=2$: the $\kappa$-independent solution ($\lambda\to0$, identical
    to balanced walls at any $\lambda$), and $\lambda=1$ with equal wall
    temperatures (equal $T$) or with equal $T/\gamma$ at the walls. The last
    two columns give, for $\lambda=1$, the Brinkman number \cref{eq:brinkman},
    with $v_{w}$ the larger wall speed, and the maximum rise of the temperature
    above $T_{w}$ for equal wall temperatures.}
    \begin{ruledtabular}
    \begin{tabular}{lccccc}
        $(\Omega_{1},\Omega_{2})$ & $\lambda\to0$ & equal $T$
        & equal $T/\gamma$ & $\Br$ & $\Delta T_{\max}/T_{w}$ \\
        \hline
        $(0.50,\,0.20)$ & $1.3737$ & $1.5538$ & $1.6463$ & $0.25$ & $0.047$ \\
        $(0.00,\,0.20)$ & $1.0480$ & $1.0236$ & $1.0792$ & $0.16$ & $3.4\times10^{-4}$ \\
        $(0.50,\,0.00)$ & $1.1551$ & $1.1872$ & $1.2623$ & $0.25$ & $0.100$ \\
        $(0.90,\,0.00)$ & $2.7121$ & $3.5280$ & $5.4167$ & $0.81$ & $1.20$ \\
    \end{tabular}
    \end{ruledtabular}
\end{table}

\subsection{Flow profiles, viscous torque, and temperature structure}
\label{subsec:profiles}

\begin{figure*}[t]
    \centering
    \includegraphics[width=0.95\textwidth]{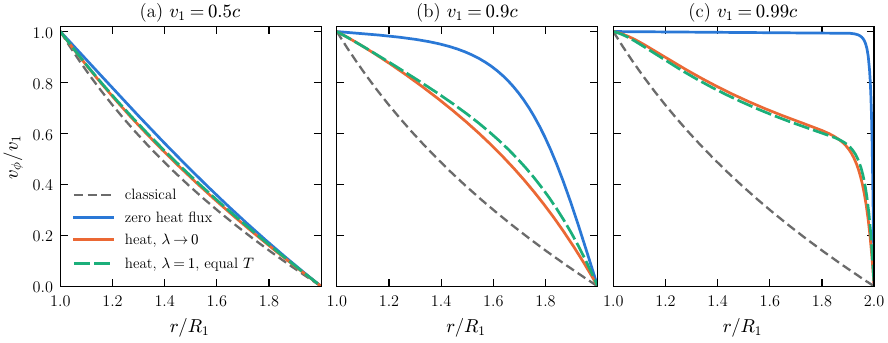}
    \caption{\label{fig:profiles}
    Velocity profiles $v_{\phi}(r)/v_{1}$ for $R_{2}/R_{1}=2$, $\Omega_{2}=0$,
    and $v_{1}=$ (a) $0.5c$, (b) $0.9c$, (c) $0.99c$: classical profile (gray
    dashed), zero-heat-flux model, and heat-conducting model for $\lambda\to0$
    (identical to balanced walls) and for $\lambda=1$ with equal wall
    temperatures.}
\end{figure*}

\begin{figure}[t]
    \centering
    \includegraphics[width=\columnwidth]{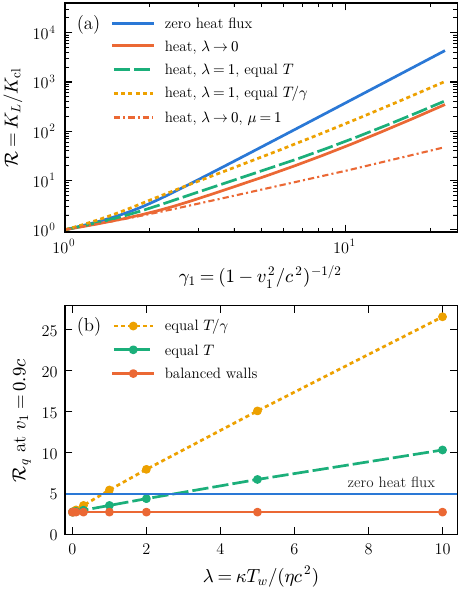}
    \caption{\label{fig:torque}
    (a) Torque ratio $K_{L}/\Kcl$ against the Lorentz factor of the inner wall
    for $R_{2}/R_{1}=2$ and $\Omega_{2}=0$. The dash-dotted line is the limit
    $\lambda\to0$ with the inertial term of Eckart's law omitted ($\mu=1$).
    The curve for equal $T/\gamma$ was computed with collocation tolerance
    $10^{-8}$ (\cref{subsec:method}).
    (b) Torque ratio of the heat-conducting model at $v_{1}=0.9c$ against
    $\lambda$ for the three thermal protocols; the horizontal line is the
    zero-heat-flux value.}
\end{figure}

\paragraph{Zero-heat-flux model.}
\Cref{tab:zero} indicates that the torque is more sensitive to the
relativistic corrections than the profile. For
$(\Omega_{1},\Omega_{2})=(0.5,0.2)$ and $(0,0.2)$, the angular velocity
differs from the classical one by less than $0.007$, whereas the torque
exceeds the classical value by $41\%$ and $5\%$, respectively, since
$|K_{L}^{(0)}|=\eta r^{3}\gamma^{3}|\Omega'|$ contains $\gamma^{3}$ explicitly.
For $\Omega_{2}=0$ the deviation grows rapidly with the wall speed. Although
the profile is still close to the classical one at $v_{1}=0.5c$
[\cref{fig:profiles}(a)], at $0.9c$ and $0.99c$ the velocity stays
close to $v_{1}$ over most of the gap and falls to zero in a layer at the outer
cylinder [\cref{fig:profiles}(b,c)]. At $v_{1}=0.99c$ the velocity drops from
$0.9v_{1}$ to $0.1v_{1}$ within about $0.02R_{1}$ of the outer wall, and the
torque ratio is $134$; at $0.999c$ it is $4.2\times10^{3}$
[\cref{fig:torque}(a)]. On the logarithmic axes of \cref{fig:torque}(a),
the local slope of $\mathcal{R}_{0}$ between $0.99c$ and $0.999c$ is
$2.99$--$3.00$, consistent with $\mathcal{R}_{0}\propto\gamma_{1}^{3}$ at large
Lorentz factors, whereas the corresponding slopes of the heat-conducting
solutions, about $2.2$--$2.4$, still increase at $0.999c$.

\paragraph{Heat-conducting model.}
In all four cases of \cref{tab:heat}, the torque ratio of the
$\kappa$-independent solution is smaller than $\mathcal{R}_{0}$, and for
$\Omega_{2}=0$ the difference grows with the wall speed: the ratio
$\mathcal{R}_{0}/\mathcal{R}_{q}$ is $1.05$ at $v_{1}=0.5c$, $1.80$ at $0.9c$,
and $5.8$ at $0.99c$. In all cases shown in \cref{fig:profiles}, the
heat-conducting profiles lie between the classical and the zero-heat-flux
profiles; at $0.99c$ they also steepen near the outer wall, although over a
wider region than the zero-heat-flux profile. Neglecting the inertial term of
Eckart's law ($\mu=1$) reduces the torque ratio from $23.2$ to $10.0$ at
$v_{1}=0.99c$ and from $339$ to $46.9$ at $0.999c$ [\cref{fig:torque}(a)].

For $\lambda=1$ the torque depends on the wall temperatures, and the sign of
the change relative to the $\kappa$-independent solution follows the rule of
\cref{subsec:thermal} in all eight combinations of case and protocol in
\cref{tab:heat}. More
specifically, $\delta$ is positive except for equal $T/\gamma$ in the case
$(0,0.2)$, where $\delta=-0.048\,T_{w}$, and $\Kcl$ is negative only for
$(0,0.2)$. The torque ratio therefore increases in all cases except $(0,0.2)$
with equal wall temperatures ($\delta=0.038\,T_{w}$), where it decreases from
$1.048$ to $1.024$. Equal $T/\gamma$ gives a larger torque than equal wall
temperatures and, in the four cases of \cref{tab:heat}, also a larger torque
than the zero-heat-flux model; at $v_{1}=0.99c$, however, the zero-heat-flux
value is the larger one ($134$ against $64.6$). At $v_{1}=0.999c$, the
$\lambda=1$ solutions give $\mathcal{R}_{q}=402$ for equal wall temperatures
and $995$ for equal $T/\gamma$ (\cref{subsec:method}).

The dependence on $\lambda$ at $v_{1}=0.9c$ is shown in
\cref{fig:torque}(b). Whereas balanced walls give $\mathcal{R}_{q}=2.71$ for
every $\lambda$, for the other two protocols the torque ratio increases
monotonically with $\lambda$, approximately linearly for $\lambda\gtrsim2$,
and reaches $10.3$ (equal $T$) and $26.6$ (equal $T/\gamma$) at $\lambda=10$.
It exceeds $\mathcal{R}_{0}=4.89$ above a value of $\lambda$ that lies between
$2$ and $5$ for equal wall temperatures and between $0.3$ and $1$ for equal
$T/\gamma$. It follows
that the zero-heat-flux model is neither an upper nor a lower bound on the
heat-conducting solutions.

The growth of the torque with $\lambda$ is accompanied by a change in the
direction of the heat flux at the inner wall. By \cref{eq:q_integrals}, the
sign of $q$ is that of $K_{E}-\Omega K_{L}$. For the $\kappa$-independent
solution, $K_{E}/K_{L}=\langle\Omega\rangle_{\mu}$ lies strictly between
$\Omega_{2}$ and $\Omega_{1}$, so that $q$ vanishes at an interior radius
where $\Omega=\langle\Omega\rangle_{\mu}$ and the viscous heat leaves through
both walls; at $v_{1}=0.9c$, $q(R_{1})=-4.90$ and $q(R_{2})=1.86$ in units of
$\eta c^{2}/R_{1}$. For $\delta>0$, however, the conductivity term of
\cref{eq:KE_exact} increases $K_{E}/K_{L}$, which at $v_{1}=0.9c$ exceeds
$\Omega_{1}$ for $\lambda>6.1$ with equal wall temperatures and for
$\lambda>1.8$ with equal $T/\gamma$. In this regime $q(R_{1})>0$, and the
inner wall supplies heat to the fluid in addition to mechanical work. Although
this reversal illustrates the effect of the conductivity term, it is a
consequence of the growth of $K_{E}$ rather than its cause, since the torque
ratio already increases with $\lambda$ at smaller values.

\begin{figure}[t]
    \centering
    \includegraphics[width=\columnwidth]{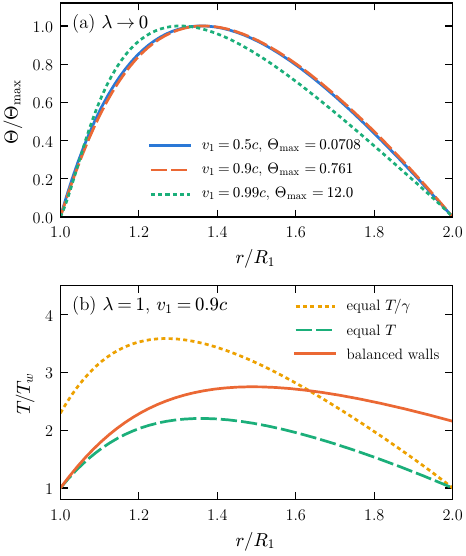}
    \caption{\label{fig:temperature}
    Temperature profiles of the heat-conducting model for $R_{2}/R_{1}=2$ and
    $\Omega_{2}=0$. (a) $\Theta=\kappa T/(\eta c^{2})$ for $\lambda\to0$,
    normalized to its maximum, at three wall speeds. (b) $T/T_{w}$ for
    $\lambda=1$ and $v_{1}=0.9c$ with the three thermal protocols.}
\end{figure}

\paragraph{Temperature.}
For $\lambda\to0$, the function $\Theta=\kappa T/(\eta c^{2})$ vanishes at both
walls and has a single maximum, at $r\simeq1.36R_{1}$ for $v_{1}=0.5c$ and
$0.9c$ and at $1.31R_{1}$ for $0.99c$ [\cref{fig:temperature}(a)]. Its maximum
value is $0.071$, $0.76$, and $12.0$ at the three speeds, so that the
temperature rise above $T_{w}$ is approximately $\eta c^{2}\Theta_{\max}/\kappa$
and diverges as $\lambda\to0$. For $\lambda=1$ and $v_{1}=0.9c$
[\cref{fig:temperature}(b)], the temperature exceeds that of the hotter wall
by at most $1.20\,T_{w}$ for equal wall temperatures, $1.29\,T_{w}$ for equal
$T/\gamma$, and $0.60\,T_{w}$ for balanced walls, for which
$T(R_{2})=2.16\,T(R_{1})$. The corresponding values are $0.10$, $0.05$, and
$0.06$ at $v_{1}=0.5c$, and $16.9$, $36.2$, and $5.6$ at $0.99c$.

The small value of $\Delta T_{\max}$ for $(\Omega_{1},\Omega_{2})=(0,0.2)$ in
\cref{tab:heat} has a different origin. In this case the temperature lies
below $T_{w}$ over $89\%$ of the gap, with a minimum of $0.993\,T_{w}$, and
exceeds $T_{w}$ by at most $3.4\times10^{-4}T_{w}$ near the inner wall. Since
the fluid rotates fastest at the outer wall, the profile $T\propto1/\mu$
favored by the inertial term increases mainly near that wall; with equal wall
temperatures, the inertial contribution therefore lowers the interior
temperature below $T_{w}$, whereas the viscous heating, which is weaker than in
the other cases, raises it above $T_{w}$ only slightly.

\subsection{Limiting cases}
\label{subsec:checks}

\begin{figure*}[t]
    \centering
    \includegraphics[width=0.95\textwidth]{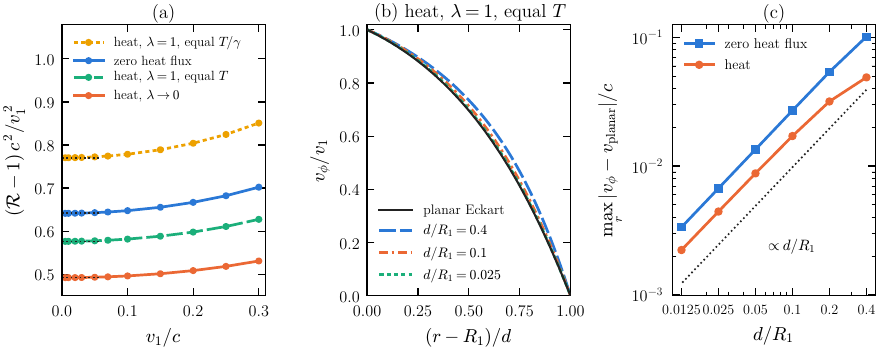}
    \caption{\label{fig:limits}
    Limiting cases for $R_{1}=1$ and $\Omega_{2}=0$. (a) Numerical torque
    correction $(\mathcal{R}-1)c^{2}/v_{1}^{2}$ for $R_{2}/R_{1}=2$ (symbols and
    lines), compared with the analytic $O(v^{2}/c^{2})$ coefficients of
    \cref{eq:pn_zero,eq:pn_heat,eq:KE0} (dotted segments at small $v_{1}$).
    (b) Narrow-gap profiles of the heat-conducting model ($\lambda=1$, equal
    wall temperatures, $v_{1}=0.9c$) for three gap widths, compared with the
    boosted planar solution of \cref{eq:planar_symmetric}. (c) Maximum
    deviation from the planar solutions for both models (heat-conducting model
    with $\lambda=1$ and equal wall temperatures); the dotted line has unit
    slope.}
\end{figure*}

The weakly relativistic and narrow-gap limits of \cref{subsec:limits} provide
two independent checks of the numerical solutions.
\Cref{fig:limits}(a) shows the torque correction
$(\mathcal{R}-1)c^{2}/v_{1}^{2}$ for small $v_{1}$ and $\Omega_{2}=0$;
extrapolation of the numerical values to $v_{1}\to0$ reproduces the
coefficients $0.6414$, $0.4921$, $0.5764$, and $0.7704$ to a relative accuracy
of $3\times10^{-7}$ or better. The limit $\lambda\to0$ at fixed wall
temperatures is approached linearly: at $v_{1}=0.9c$ with equal wall
temperatures,
$\mathcal{R}_{q}=2.71218$, $2.71289$, and $2.71996$ for $\lambda=10^{-4}$,
$10^{-3}$, and $10^{-2}$, compared with $2.71211$ for $\lambda\to0$.

In the narrow-gap limit [\cref{fig:limits}(b,c)], the cylindrical solutions
for $v_{1}=0.9c$ and $\Omega_{2}=0$ approach the planar profiles as $d/R_{1}$
decreases. The maximum deviation decreases from $0.049c$ at $d/R_{1}=0.4$ to
$2.2\times10^{-3}c$ at $d/R_{1}=0.0125$ for the heat-conducting model
($\lambda=1$, equal wall temperatures), and from $0.10c$ to
$3.4\times10^{-3}c$ for the zero-heat-flux model; over the three narrowest gaps
it is proportional to $d/R_{1}$ (\cref{tab:checks}), as expected for a
correction of relative order $d/R_{1}$.

\section{Discussion and conclusions}
\label{sec:discussion}

\subsection{Physical context and scope}
\label{subsec:context}

The flow considered here is not a model of a specific astrophysical system,
and its relevance is of two kinds. First, it is a stationary solution of
relativistic dissipative hydrodynamics in which angular-momentum and energy
transport can be computed to high accuracy, so that it can serve as a test
problem for numerical codes. Second, it isolates three effects that are
entangled in realistic rotating relativistic flows: the kinematic Lorentz
factors of a sheared rotation, the momentum carried by heat, and the inertial
(Tolman) contribution to heat conduction in a rotating fluid. Although these
effects are expected to matter in systems such as the inner regions of
accretion flows, differentially rotating compact stars, and rotating
relativistic plasmas, such systems also involve gravity, magnetic fields, and
turbulence, which are absent here.

\subsection{Limitations and extensions}
\label{subsec:limitations}

The analysis is restricted to flat spacetime, stationary and laminar
axisymmetric flow, infinitely long cylinders, a fluid with a conserved particle
number, and constant transport coefficients. As discussed in
\cref{subsec:thermal}, the last assumption is the most restrictive at large
wall speeds, since for the thermal protocols considered the temperature varies
by about $40\%$ or more at $v_{1}\gtrsim0.9c$; quantitative
statements at these speeds therefore require temperature-dependent
coefficients. In addition, for $\Omega_{2}=0$ the laminar base flow is
centrifugally unstable above a critical Taylor number (\cref{subsec:cases}).

Independently of the temperature dependence of $\eta$ and $\kappa$, the
first-order description requires the dissipative fluxes to be small compared
with the enthalpy density, $|\pi^{\hat{r}\hat{\phi}}|/(e+P)\ll1$ and
$|q|/[(e+P)c]\ll1$, where $\pi^{\hat{r}\hat{\phi}}=-\eta r\gamma^{2}\Omega'$ is
the shear stress in the local rest frame. Since $e$ and $P$ do not enter
\cref{eq:reduced,eq:temperature}, both ratios are proportional to the inverse
Reynolds number $\chi\equiv\eta c/[(e+P)R_{1}]$, with coefficients fixed by
the dimensionless solution. For $R_{2}/R_{1}=2$ and $\Omega_{2}=0$, the largest
values of $r\gamma^{2}|\Omega'|R_{1}/c$ and $|q|R_{1}/(\eta c^{2})$ in the gap
are at most about $7$ and $5$ at $v_{1}=0.9c$ and about $62$ and $51$ at
$0.99c$, both for $\lambda\to0$ and for $\lambda=1$ with equal wall
temperatures, so that the first-order description requires $\chi\ll0.1$ and
$\chi\ll0.02$, respectively. Such values can be
chosen at any fixed $v_{1}<c$. However, a small $\chi$ corresponds to a large
Reynolds number, $\mathrm{Re}=(v_{1}/c)/\chi$, so that for $\Omega_{2}=0$ this
requirement competes with the stability condition of \cref{subsec:cases}.

The analysis is based on first-order (Eckart) hydrodynamics, which is acausal
and whose equilibria are unstable to linear perturbations
\cite{HiscockLindblom1985}. These properties, however, concern the evolution
of perturbations rather than the existence of a stationary solution. In causal
theories of Israel--Stewart type \cite{IsraelStewart1979}, the dissipative
fluxes relax, on a time scale set by the microscopic interactions, to states
that are essentially indistinguishable from those of the first-order theory
\cite{Geroch1995,Lindblom1996}; the stationary base flow obtained here is
therefore expected to agree with that of causal theories to leading order in
gradients. Second-order theories also contain terms such as the comoving
derivative of the shear stress and its coupling to the vorticity
\cite{Baier2008,Denicol2012}, which need not vanish on the curved streamlines
of the present flow. Corrections of relative order of the ratio of the mean
free path to the gap width are therefore to be expected.

Natural extensions of the present work are temperature-dependent transport coefficients, the uncharged
fluid in the Landau frame
\cite{LandauLifshitz1987,Gavassino2026,GavassinoSchlichtingDenicol2026} or in
causal first-order
(BDNK) frames \cite{BemficaDisconziNoronha2022,Kovtun2019,HoultKovtun2020}, the linear stability of the
base flow, and curved-spacetime backgrounds.

\subsection{Conclusions}
\label{subsec:conclusions}

The comparison of the two formulations establishes that the zero-heat-flux
model is not a consistent description of a stationary state. Its energy
equation is violated by the local heating rate divided by $\gamma$, and its
torque differs from that of the $\kappa$-independent energy-conserving
solution by a factor that
grows with the wall speed; for $R_{2}/R_{1}=2$ and a resting outer cylinder,
this factor is $1.80$ at $v_{1}=0.9c$ and $5.8$ at $0.99c$.

In the heat-conducting formulation, the exact energy-flux relation
\cref{eq:KE_exact} reduces the influence of the thermal boundary conditions to
the single mismatch $\delta$. Consequently, the $\kappa$-independent solution,
which in the planar problem is associated with equal plate temperatures, is
associated here with balanced walls; it coincides with the limit $\kappa\to0$
at fixed wall temperatures, in which, in contrast to the planar case, the
inertial term of Eckart's law survives through the weight $\mu$. Since
$\mu(R_{2})$ depends on the profile, the balance condition is a property of
the solution rather than an independent choice of thermostat temperatures,
and balanced walls require the wall temperatures to be adjusted to the flow.
The importance of the inertial term increases with the Lorentz factor: omitting it changes
the torque by $10\%$ at $0.9c$ and by a factor of $2.3$ at $0.99c$. For other
wall temperatures, the torque depends on $\lambda$ and on the protocol, and
with the outer cylinder at rest it grows approximately linearly with
$\lambda$, which indicates that, within the present model, the torque cannot
be specified without the thermal boundary conditions.

These results are subject to the assumption of constant transport
coefficients, which becomes restrictive at $v_{1}\gtrsim0.9c$, where the
temperature varies by about $40\%$ or more. Within this
restriction, the base flow provides a test problem with known first integrals,
an exact energy-flux relation, a $\kappa$-independent family of solutions,
and analytic weakly relativistic and narrow-gap limits, for which benchmark
values are listed in \cref{app:numerics}.

\onecolumngrid
\begin{acknowledgments}
The author thanks Prof.~Andria Rogava for guidance and consultation during the
development of this work, and Ilia State University, where this research was
conducted during the author's employment as a research assistant.
\end{acknowledgments}

\section*{Data availability}
The code used for the numerical results, the computed data, and the scripts
that produce all figures and tables of this paper are openly available in a
GitHub repository~\cite{GelekvaCode2026}:
\begin{center}
\mbox{\footnotesize\href{https://github.com/mean-ani-mustard/relativistic-taylor-couette}{\texttt{github.com/mean-ani-mustard/relativistic-taylor-couette}}}
\end{center}
\clearpage
\twocolumngrid

\appendix

\section{Divergence identities in cylindrical coordinates}
\label{app:identities}

For a symmetric tensor $T^{\alpha\beta}$ that depends only on $r$, the
covariant divergence is
$\cd_{\alpha}T^{\alpha\beta}=\pd_{\alpha}T^{\alpha\beta}
+\Gamma^{\alpha}_{\alpha\rho}T^{\rho\beta}
+\Gamma^{\beta}_{\alpha\rho}T^{\alpha\rho}$, with
$\Gamma^{\alpha}_{\alpha r}=1/r$ the only nonzero contraction. Using
\cref{eq:christoffel}, the three nontrivial components are
\begin{align}
    \cd_{\alpha}T^{\alpha t} &= \frac{1}{r}\frac{\dd}{\dd r}\bigl(rT^{rt}\bigr),
    \label{eq:div_t}\\
    \cd_{\alpha}T^{\alpha\phi} &= \frac{\dd T^{r\phi}}{\dd r} + \frac{3}{r}T^{r\phi}
    = \frac{1}{r^{3}}\frac{\dd}{\dd r}\bigl(r^{3}T^{r\phi}\bigr),
    \label{eq:div_phi}\\
    \cd_{\alpha}T^{\alpha r} &= \frac{1}{r}\frac{\dd}{\dd r}\bigl(rT^{rr}\bigr)
    - rT^{\phi\phi}.
    \label{eq:div_r}
\end{align}
In \cref{eq:div_phi}, $1/r$ comes from $\Gamma^{\alpha}_{\alpha r}$ and $2/r$
from $\Gamma^{\phi}_{r\phi}+\Gamma^{\phi}_{\phi r}$; in \cref{eq:div_r}, the
last term comes from $\Gamma^{r}_{\phi\phi}=-r$. The $z$ component vanishes
identically.

\section{Components of the shear tensor}
\label{app:shear}

The covariant derivatives entering \cref{eq:shear_def} are
$\cd_{r}U_{\phi}=\pd_{r}(r^{2}\gamma\Omega)-\Gamma^{\phi}_{r\phi}U_{\phi}
=r\gamma\Omega+r^{2}(\gamma\Omega)'$ and
$\cd_{\phi}U_{r}=-\Gamma^{\phi}_{\phi r}U_{\phi}=-r\gamma\Omega$, so that
\begin{equation}
    \cd_{(r}U_{\phi)} = \tfrac12 r^{2}(\gamma\Omega)',
    \qquad
    \cd_{(r}U_{t)} = -\tfrac12\gamma',
    \label{eq:sym_derivs}
\end{equation}
where the second relation follows from $U_{t}=-\gamma$ and the absence of
Christoffel symbols with indices $r$ and $t$. Because $U_{r}=0$, one has
$h_{r}{}^{\alpha}=\delta_{r}{}^{\alpha}$, and the projector components entering
the remaining contractions are
\begin{equation}
\begin{aligned}
    h_{\phi}{}^{\phi} &= \gamma^{2}, &\quad
    h_{\phi}{}^{t} &= r^{2}\gamma^{2}\Omega,\\
    h_{t}{}^{t} &= -r^{2}\gamma^{2}\Omega^{2}, &\quad
    h_{t}{}^{\phi} &= -\gamma^{2}\Omega .
\end{aligned}
    \label{eq:projector}
\end{equation}
Combining \cref{eq:sym_derivs,eq:projector} with
$(\gamma\Omega)'=\gamma'\Omega+\gamma\Omega'$ gives
\begin{equation}
\begin{aligned}
    \sigma_{r\phi} &= \tfrac12 r^{2}\gamma^{2}\bigl[(\gamma\Omega)'-\Omega\gamma'\bigr]
    = \tfrac12 r^{2}\gamma^{3}\Omega',\\
    \sigma_{rt} &= -\tfrac12 r^{2}\gamma^{3}\Omega\Omega' ,
\end{aligned}
    \label{eq:shear_lower}
\end{equation}
and raising the indices with $g^{rr}=1$, $g^{\phi\phi}=r^{-2}$, and $g^{tt}=-1$
yields \cref{eq:shear_component}. All other components vanish. Contracting,
$\sigma_{\mu\nu}\sigma^{\mu\nu}=2\bigl(\sigma_{r\phi}\sigma^{r\phi}
+\sigma_{rt}\sigma^{rt}\bigr)=\tfrac12 r^{2}\gamma^{4}\Omega'^{2}$.

\section{Numerical verification and benchmark values}
\label{app:numerics}

\Cref{tab:checks} summarizes the accuracy checks of the numerical solutions,
and \cref{tab:benchmark} lists the flux constants for the cases of
\cref{tab:zero,tab:heat} ($R_{1}=1$, $R_{2}=2$, $c=\eta=T_{w}=1$, so that
$\kappa=\lambda$) for use as a test problem. The heat-conducting values were
obtained independently by collocation and by two-parameter shooting.

\begin{table}[t]
    \caption{\label{tab:checks}
    Accuracy checks. Relative deviations are maxima over the cases of
    \cref{tab:zero,tab:heat} and all thermal settings unless stated otherwise.}
    \begin{ruledtabular}
    \begin{tabular}{lc}
        Check & Largest deviation \\
        \hline
        First integrals \cref{eq:first_integrals}, $4001$ points & $1.4\times10^{-7}$ \\
        Energy budget \cref{eq:budget} & $8\times10^{-16}$ \\
        Local energy balance, integrated & $2.5\times10^{-12}$ \\
        Energy-flux relation \cref{eq:KE_exact} & $2.7\times10^{-11}$ \\
        Shooting vs.\ collocation, $(K_{L},K_{E})$ & $5.4\times10^{-11}$ \\
        Balanced walls vs.\ $\lambda\to0$\footnotemark[4] & $6.7\times10^{-12}$ \\
        Collocation tolerance $10^{-8}\to10^{-9}$, $\mathcal{R}_{q}$\footnotemark[1]
            & $6.8\times10^{-12}$ \\
        Continuation steps doubled, $\mathcal{R}_{q}$\footnotemark[2]
            & $2\times10^{-16}$ \\
        ODE tolerance $10^{-12}\to10^{-13}$, $K_{L}^{(0)}$\footnotemark[3]
            & $4.7\times10^{-13}$ \\
        $O(v^{2}/c^{2})$ coefficients, \cref{subsec:limits} & $2.8\times10^{-7}$ \\
        Narrow gap: slope vs.\ $d/R_{1}$\footnotemark[5] & $0.99$, $1.00$ \\
        Continuation in $v_{1}$ vs.\ in $T(R_{1})$\footnotemark[6] & $1.1\times10^{-11}$ \\
    \end{tabular}
    \end{ruledtabular}
    \footnotetext[1]{At $v_{1}=0.9c$ and $0.99c$; from $10^{-6}$ to $10^{-9}$
    the change is at most $4.3\times10^{-9}$.}
    \footnotetext[2]{At $v_{1}=0.99c$ ($\lambda=1$) and $0.999c$ ($\lambda\to0$).}
    \footnotetext[3]{At $v_{1}=0.5c$, $0.9c$, and $0.99c$.}
    \footnotetext[4]{For $\lambda=1$ in all cases of \cref{tab:heat}, and for
    $\Omega_{2}=0$ with $\lambda\leq20$ at $v_{1}=0.5c$ and $0.9c$ and
    $\lambda\leq5$ at $0.99c$.}
    \footnotetext[5]{Log-log slope of the maximum deviation from the planar
    solution over $d/R_{1}=0.05$, $0.025$, and $0.0125$ at $v_{1}=0.9c$,
    for the heat-conducting and the zero-heat-flux model, respectively.}
    \footnotetext[6]{Equal $T/\gamma$, $\lambda=1$, $v_{1}=0.995c$--$0.999c$.}
\end{table}

\begin{table}[t]
    \caption{\label{tab:benchmark}
    Angular-momentum flux $K_{L}$ (for the zero-heat-flux model,
    $K_{L}^{(0)}$) and energy flux $K_{E}$ for $R_{1}=1$, $R_{2}=2$, and
    $c=\eta=T_{w}=1$. Balanced walls give the $\lambda\to0$ values for every
    $\lambda$.}
    \begin{ruledtabular}
    \begin{tabular}{lrr}
        Model & $K_{L}$ & $K_{E}$ \\
        \hline
        \multicolumn{3}{l}{$(\Omega_{1},\Omega_{2})=(0.50,\,0.20)$} \\
        zero heat flux & $1.1251230794$ & --- \\
        $\lambda\to0$ & $1.0989964156$ & $0.3524620174$ \\
        $\lambda=1$, equal $T$ & $1.2430579693$ & $0.6040617562$ \\
        $\lambda=1$, equal $T/\gamma$ & $1.3170769785$ & $0.7343134427$ \\
        \hline
        \multicolumn{3}{l}{$(\Omega_{1},\Omega_{2})=(0.00,\,0.20)$} \\
        zero heat flux & $-0.5624246334$ & --- \\
        $\lambda\to0$ & $-0.5589162434$ & $-0.0702574253$ \\
        $\lambda=1$, equal $T$ & $-0.5459249467$ & $-0.0137027569$ \\
        $\lambda=1$, equal $T/\gamma$ & $-0.5755831824$ & $-0.1413348957$ \\
        \hline
        \multicolumn{3}{l}{$(\Omega_{1},\Omega_{2})=(0.50,\,0.00)$} \\
        zero heat flux & $1.6163281523$ & --- \\
        $\lambda\to0$ & $1.5401690424$ & $0.3274937446$ \\
        $\lambda=1$, equal $T$ & $1.5829515014$ & $0.4376997796$ \\
        $\lambda=1$, equal $T/\gamma$ & $1.6830007406$ & $0.6906176419$ \\
        \hline
        \multicolumn{3}{l}{$(\Omega_{1},\Omega_{2})=(0.90,\,0.00)$} \\
        zero heat flux & $11.7372012758$ & --- \\
        $\lambda\to0$ & $6.5090520533$ & $3.7209313138$ \\
        $\lambda=1$, equal $T$ & $8.4672519975$ & $5.8613934413$ \\
        $\lambda=1$, equal $T/\gamma$ & $13.0000912188$ & $10.7664903291$ \\
    \end{tabular}
    \end{ruledtabular}
\end{table}

\FloatBarrier

\bibliography{references}

\end{document}